\documentclass[epsf]{aa}
\usepackage{natbib}
\usepackage{epsfig}

\newcommand{\kms}{\mbox{km\,s$^{-1}$}}

\def\ms{\,m\,s$^{-1}$}         
\def\kms{\,km\,s$^{-1}$}       

\def\kms{km\, s$^{-1}$}

\begin{document}
\title{VLT transit and occultation photometry for the bloated planet CoRoT-1b\thanks{Based on data collected with the VLT/FORS2 and VLT/HAWK-I instruments at ESO Paranal Observatory, Chile (programs 080.C-0661(B) and 382.C-0642(A)).}$^{, }$\thanks{The photometric time-series used in this work are only available in electronic form at the CDS via anonymous ftp to  cdsarc.u-strasbg.fr (130.79.128.5) or via http://cdsweb.u-strasbg.fr/cgi-bin/qcat?J/A+A/}}
\author{M. Gillon$^{1,2}$, B.-O. Demory$^{2}$, A.~H.~M.~J. Triaud$^{2}$, T. Barman$^3$, L. Hebb$^{4}$, J. Montalb\'an$^1$,    P.~F.~L. Maxted$^{5}$, D. Queloz$^{2}$, M. Deleuil$^6$, P. Magain$^{1}$}   

\offprints{michael.gillon@obs.unige.ch}
\institute{
$^1$ Institut d'Astrophysique et de G\'eophysique,  Universit\'e
  de Li\`ege,  17 All\'ee du 6 Ao\^ut,  Bat.  B5C, Li\`ege 1, Belgium \\
$^2$  Observatoire de Gen\`eve, Universit\'e de Gen\`eve, 51 Chemin des Maillettes, 1290 Sauverny, Switzerland\\
$^3$ Lowell Observatory, 1400 West Mars Hill Road, Flagstaff, AZ 86001, USA\\
$^4$ School of Physics and Astronomy, University of St. Andrews, North Haugh, Fife, KY16 9SS, UK\\
$^5$ Astrophysics Group, Keele University, Staffordshire, ST5 5BG, UK\\
$^6$ Laboratoire d'Astrophysique de Marseille (UMR 6110), 38  rue Fr\'ed\'eric Joliot-Curie, 13388 Marseille, France\\
}

\date{Received date / accepted date}
\authorrunning{M. Gillon et al.}
\titlerunning{VLT transit and occultation photometry for CoRoT-1b}
\abstract{We present VLT eclipse photometry for the  giant planet CoRoT-1b. We observed a transit in the R-band filter and an occultation in a narrow filter centered on 2.09 $\mu$m.  Our analysis of this new photometry and published radial velocities, in combination with stellar-evolutionary modeling, leads to a planetary mass and radius of  1.07$^{+0.13}_{-0.18} M_{Jup}$ and 1.45$^{+0.07}_{-0.13} R_{Jup}$, confirming the very low density previously deduced from CoRoT photometry. The large occultation depth that we measure at 2.09 $\mu$m (0.278$^{+0.043}_{-0.066} \%$) is consistent with thermal emission and is better reproduced by an atmospheric model with no redistribution of the absorbed stellar flux to the night side of the planet.
 }

\maketitle

\section{Introduction}

Transiting planets play an important role in the study of planetary objects outside    our 
solar system. Not only can we                infer their density and use it to constraint      their
composition, but several other interesting measurements are possible for these objects 
(see e.g. review by Charbonneau et al. 2007). In particular,  their thermal emission  can                
be measured during their occultation, allowing the study of their   atmosphere     without  
spatially resolving their light from that of the host star. The                    $Spitzer$ $Space$      
$Telescope$ (Werner et al. 2004) has produced a flurry of such planetary          emission 
measurements, all at wavelengths longer than 3.5 $\mu$m.  From the ground,     several 
attempts to obtain  occultation measurements at shorter wavelengths than        $Spitzer$ 
spectral window were performed (Richardson et al. 2003a,b;  Snellen 2005; Deming   et  
al. 2007; Knutson et al. 2007, Snellen \& Covino 2007, Winn et al. 2008).    Very recently, 
two of them were successful:  Sing \& L\'opez-Morales (2009) obtained a $\sim$              4 
$\sigma$ detection of the occultation of OGLE-TR-56b in the $z$-band       (0.9 $\mu$m) , 
while De Moiij \& Snellen (2009) detected at  $\sim$ 6 $\sigma$ the thermal   emission of 
TrES-3b in the K-band (2.2 $\mu$m). It is important to obtain more similar measurements 
to improve our understanding of the atmospheric properties of short-period        extrasolar 
planets.
   
CoRoT-1b (Barge et al. 2008, hereafter B08) was the first planet detected by the  CoRoT 
space transit survey (Baglin et al. 2006). With an orbital period of 1.5 days,                   this 
Jupiter-mass planet orbits at only $\sim$ 5 stellar radii from its G0V host star. Due to  this 
proximity, its stellar irradiation is clearly large enough ($\sim$ $3.9 \times 10^{9}$ erg~s$^{-1}$~cm$^{-2}$) to make it join OGLE-TR-56b, TrES-3b and a few other planets within the pM theoretical class proposed by Fortney et al. (2008). Under this theory, pM planets receive a stellar flux large enough to have high-opacity compounds like TiO and VO present in their gaseous form in the day-side atmosphere. These compounds should be responsible for a stratospheric thermal inversion, with re-emission on a very short time-scale of a large fraction of the incoming stellar flux, resulting in a poor efficiency of the  heat distribution from the day-side to the night-side and to  enhanced infrared planetary fluxes at orbital phases close to the occultation. Like the other pM planets, CoRoT-1b is  thus a  good target for near-infrared occultation measurements. Furthermore, CoRoT-1b belongs to the subgroup of the planets with a radius larger than predicted by basic models of irradiated planets (e.g. Burrows et al. 2007, Fortney et al. 2007).  Tidal heating has been proposed by several authors (e.g. Bodenheimer et al. 2001, Jackson et al. 2008b) as a possible extra source of energy able to explain the radius anomaly shown by these hyper-bloated planets.  As shown by Jackson et al. (2008b) and Ibgui \& Burrows (2009), even a tiny orbital eccentricity  is able to produce an intense tidal heating for very short period planets. Occultation photometry does not only allow to measure the planetary thermal emission, but also constrains strongly the orbital eccentricity (see e.g. Charbonneau et al. 2005). Such an occultation measurement for CoRoT-1b could thus help understanding its low density. 

These reasons motivated us to measure  an occultation of CoRoT-1b with the Very Large Telescope (VLT). We also decided to obtain a precise VLT transit light curve for this planet to better constrain its orbital elements. Furthermore, CoRoT transit photometry presented in B08 is exquisite, but it is important to obtain an independent measurement of similar quality to check its reliability and to assess the presence of  any systematic effect in the CoRoT photometry.

We present in Section 2 our new VLT data and their reduction. Section 3 presents our analysis of the resulting photometry and our determination of the system parameters. Our results are discussed in Section 4, before giving our conclusion in Section 5.
  
\section{Observations}

\subsection{VLT/FORS2 transit photometry}

A transit of CoRoT-1b was observed on 2008 February 28 with the FORS2 camera (Appenzeller et al. 1998) installed at the VLT/UT1 (Antu). FORS2 camera has a mosaic of two 2k $\times$ 4k MIT 
CCDs and is optimized for observations in the red with a very low level of fringes. It was used several times in the past to obtain high precision transit photometry (e.g. Gillon et al. 2007a, 2008). The high resolution mode was used to optimize the spatial sampling, resulting in a 4.6' $\times$ 4.6' field of view with a  pixel scale of  0.063''/pixel. Airmass increased from 1.08 to 1.77 during the run which lasted  from 1h16 to 4h30 UT. The quality of the night was photometric. Due to scheduling constraints,  only a small amount of observations were performed before and after the transit, and the total out-of-transit (OOT) part of the run is only $\sim$ 50 minutes. 

114 images were acquired  in the R\_SPECIAL filter ($\lambda_{eff}= 655 $ nm, FWHM = 165 nm) with an exposure time of 15~s. After a standard pre-reduction,  the stellar fluxes were extracted for all the images with the {\tt IRAF}\footnote{ {\tt IRAF} is distributed by the National Optical Astronomy Observatory, which is operated by the Association of Universities for Research in Astronomy, Inc., under cooperative agreement with the National Science Foundation.} {\tt DAOPHOT}  aperture photometry software (Stetson, 1987). We noticed that CoRoT-1 was saturated in 11 images because of seeing and transparency variations, and we rejected these images from our analysis. Several sets of reduction parameters were tested, and we kept the one giving the most precise photometry for the stars of similar brightness than CoRoT-1. After a careful selection of reference stars, differential photometry was obtained.  
A linear fit for magnitude $vs$ airmass  was performed to correct the photometry for differential reddening using the OOT data. The corresponding fluxes were then normalized using the OOT part of the photometry. The resulting transit light curve is shown in Fig. 1. After subtraction of the best-fit model (see next section), the obtained residuals show a $rms$ of $\sim$ 520 ppm, very close to the photon noise limit ($\sim$ 450 ppm). 

\begin{figure}
\label{fig:a}
\centering                     
\includegraphics[width=9cm]{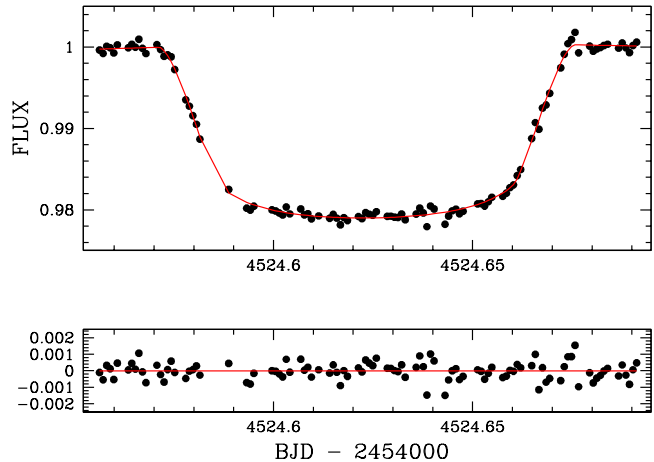}
\caption{$Top$: VLT/FORS2 R-band transit light curve with the best-fitting transit + trend model superimposed. $Bottom$: residuals of the fit.}
\end{figure}

\subsection{VLT/HAWK-I occultation photometry}

We observed an occultation of CoRoT-1b with HAWK-I ({\it High Acuity
Wide-field K-band Imager}, Pirard et al. 2004, Casali et al. 2006), a cryogenic near-IR
imager recently installed at the VLT/UT4 (Yepun). HAWK-I provides a
relatively large field of view of 7.5'  x 7.5'. The detector is kept at 75 K and is 
composed of a mosaic of four Hawaii-2RG 2048x2048 pixels chips. The pixel
scale is 0.106"/pixel, providing a good spatial sampling even for the excellent 
seeing conditions of Paranal (seeing down to 0.3 arcsec measured in K-band).

Instead of using a broad band K or K$_s$ filter, we choose to observe with the narrow band 
filter NB2090 (central wavelength =  2.095 $\mu$m, width = 0.020 $\mu$m). This filter
 avoids absorption bands at the  edge of the K-band, its small width minimizes the effect 
 of differential extinction and furthermore its bandpass shows a much  smaller sky 
 emission than the one of the  nearby Br$\gamma$ bandpass (central wavelength = 
 2.165 $\mu$m, width = 0.030 $\mu$m), leading to a flux ratio background/star more 
 than twice better than in Br$\gamma$ or K-band filters. Because of the large aperture
of the VLT and the relative brightness of CoRoT-1, the expected stellar count in 
this narrow filter is still good enough to allow theoretical noise of less than 0.15\% for a
1 minute integration.  

Observations took place on 2009 January 06 from 1h54 to 7h56 UT. Atmospheric
conditions were very good, while the mean seeing measured on the images was 0.47".  
Airmass decreased from 1.36 to 1.08 then raised to 1.65. Each exposure was composed of
4 integrations of 11~s each. A random jitter pattern within a square 45"-sized box was applied 
to the telescope. This strategy aimed to obtain for each image an accurate sky map from the 
neighbor images. Indeed, the near-IR background  shows a large spatial variability at 
different  scales and an accurate subtraction of this complex background is crucial, 
except when this background has an amplitude negligible when compared to the stellar count 
(see e.g. Alonso et al. 2008). In total, 318 images were obtained during the run.

After a standard pre-reduction (dark subtraction, flatfield division), a sky map was constructed 
and removed for each image using a median-filtered set of the ten adjacent images. The 
resulting sky-subtracted images were aligned and then compared on a per-pixel basis
to the median of the 10 adjacent images in order to detect any spurious values due, e.g., to a cosmic hit or a pixel  damage. The concerned pixels had their value replaced by  the one obtained by linear interpolation using the 10 adjacent images.

Two different methods were tested to extract the stellar fluxes.  Aperture photometry was obtained using the {\tt IRAF DAOPHOT} software and compared to deconvolution photometry obtained  with the algorithm {\tt DECPHOT} (Gillon et al. 2006 and 2007b, Magain et al. 2007). We obtained a significantly ($\sim$ 25\%) better result with  {\tt DECPHOT}. We attribute this improvement to the fact that {\tt DECPHOT}  optimizes the separation of the stellar flux from the background contribution, while aperture photometry simply sums the counts within an aperture.

In order to avoid any systematic noise due to the different characteristics of  the HAWK-I chips, we choose to use only reference stars located in the same chip than our target to obtain the differential photometry. As CoRoT-1 lies in a dense field of the Galactic plane, we have in any case enough reference flux in one single chip to reach the desired photometric precision. After a careful selection of the reference stars, the obtained differential curve shows clearly an eclipse with the expected duration and timing (Fig. 2). We could not find any firm correlation of the OOT photometric values with the airmass or time, so we simply normalized the fluxes using the OOT part without any further correction. The OOT $rms$ is 0.32 \%, much larger than the mean theoretical error: 0.13 \%. This difference implies the existence of an extra source of noise of $\sim$ 0.3 \%. We attribute this noise to the sensitivity and cosmetic inhomogeneity of the detector combined with our jitter strategy. In the optical, one can avoid this noise by staring at the same exact position during the whole run, i.e. by keeping the stars on the same pixels. In the near-IR, dithering is needed to remove properly the large, complex and variable background. This background varies in time at frequencies similar to the one of the transit, so any poor background removal is able to bring correlated noise in the resulting photometry. It is thus much preferable to optimize the background subtraction by using a fast random jitter pattern even if this brings an extra noise, because this latter is dominated by frequencies much larger than the one of the searched signal and is thus unable to produce a fake detection or modify the eclipse shape. 

\begin{figure}
\label{fig:b}
\centering                     
\includegraphics[width=9cm]{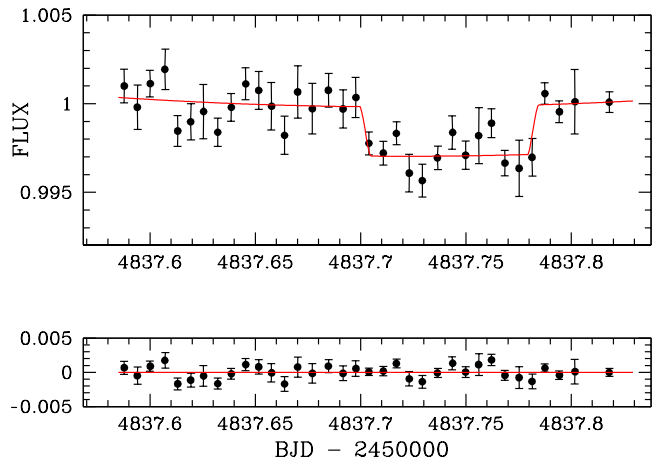}
\caption{$Top$: VLT/HAWK-I 2.09 $\mu$m occultation light curve binned per 10 minutes, with the best-fitting occultation + trend model superimposed. $Bottom$: residuals of the fit.}
\end{figure}

\section{Analysis}

\subsection{Data and model}

To obtain an independent  determination of the system parameters, we decided to use only our VLT R-band transit and 2.09 $\mu$m occultation photometry in addition to the SOPHIE (Bouchy et al. 2006) radial velocities (RV) presented in B08 as data for our analysis. 

These data were used as input into a Markov Chain Monte Carlo (MCMC; see e.g. Tegmark 2004, Gregory 2005, Ford 2006) code. MCMC is a Bayesian inference method based on stochastic simulations and provides the $a$ $posteriori$ probability distribution of adjusted parameters for a given model. Here the model is based on a star and a transiting planet on a Keplerian orbit about their center of mass. More specifically, we used a classical Keplerian model for the RV variations and we fitted independent offsets for the two epochs of the SOPHIE observations to account for the drift between them mentioned in  B08. To fit the VLT photometry, we used the photometric eclipse model of Mandel \& Agol (2002) multiplied by a trend model. In order to obtain reliable error bars for our fitted parameters, it is indeed preferable to consider the possible presence of a low-amplitude time-dependant systematic in our photometry due, e.g. to an imperfect differential extinction correction or a low-amplitude low-frequency stellar variability. We choose to model this trend as a second-order time polynomial function for both FORS2 and HAWK-I photometry.

\subsection{Limb-darkening}

For the transit, a quadratic limb darkening law was assumed, with initial coefficients $u_1$ and $u_2$ interpolated from Claret's tables (2000; 2004) for the R-band photometric filter and for $T_{eff} = 5950 \pm 150$ K, log $g$ = $4.25 \pm 0.30$ and [Fe/H] = $-0.30 \pm 0.25$ (B08).  We used the partial derivatives of $u_1$ and $u_2$ as a function of the spectroscopic parameters in Claret's tables to obtain their errors $\sigma_{u_1}$ and $\sigma_{u_2}$ via:

\begin{equation}\label{eq:1}
\sigma_{u_x}= \sqrt{\sum_{i=1}^3\,\big(\frac{\delta u_x}{\delta S_i}\sigma_{S_i}\big)^2 }\textrm{,}
 \end{equation}\noindent
 where $x$ is 1 or 2, while $S_i$ and $\sigma_{S_i}$ are the $i^{th}$ ($i=1,3$) spectroscopic parameter and its error from B08. We obtained $u_1 = 0.279 \pm 0.033$ and $u_2 = 0.351 \pm 0.016$  as initial values. We allowed $u_1$ and $u_2$ to float in our MCMC analysis, using as jump parameters not these coefficients themselves but the combinations $c_1 = 2 \times u_1 + u_2$ and $c_2 = u_1 - 2 \times u_2$ to minimize the correlation of  the  obtained uncertainties (Holman et al. 2006).  The following Bayesian prior on $c_1$ and $c_2$  was added to our merit function:
 \begin{equation}\label{eq:2}
BP_{\rm   limb-darkening} = \sum_{i=1,2} \bigg(\frac{c_i - c'_i}{\sigma_{c'_i}} \bigg)^2
 \end{equation}\noindent where $c'_i$ is the initial value deduced for the coefficient $c_i$ and $\sigma_{c'_i}$ is its error computed from $\sigma_{u_1}$ and $\sigma_{u_2}$. We let $c_1$ and $c_2$ be free parameters under the control of a Bayesian prior to propagate the uncertainty on the limb-darkening to the deduced transit parameters. 
 
 \subsection{Jump parameters}
 
The other jump parameters in our MCMC simulation were:  the transit timing (time of minimum light) $T_0$, the planet/star area ratio  $(R_p/R_s)^2 $, the transit width (from first to last contact) $W$, the impact parameter $b'=a\cos{i}/R_\ast$, three coefficients per photometric time-series for the low-frequency systematic, one systemic RV for each of the two SOPHIE epochs, and the two parameters  $e\cos{\omega}$ and $e\sin{\omega}$ where $e$ is the orbital eccentricity and $\omega$ is the argument of periastron. The RV orbital semi-amplitude $K$ was not used as jump parameter, but instead we used the following parameter:
\begin{equation}\label{eq:3}
K_2 = K  \sqrt{1-e^2}   \textrm{ } P^{1/3} = (2\pi G)^{1/3} \frac{M_p \sin i}{(M_p + M_\ast)^{2/3}}\textrm{,}
 \end{equation}\noindent to minimize the correlation with the other jump parameters.  
 
We notice that our used jump parameter $b'$ is equal to the actual transit impact parameter $b$ only for a circular orbit. For a non-zero eccentricity, it is related to the actual impact parameter $b$ via:
 \begin{equation}\label{eq:4}
b = b'  \textrm{ } \frac{1 - e^2}{1+ e \sin{\omega}}\textrm{.}
 \end{equation}\noindent Here too, the goal of using $b'$ instead of $b$ is to minimize the correlation between the jump parameters.  
 
The orbital period $P$ was let free in our analysis, constrained not only with the data presented above but also with the timings determined independently by Bean (2009) for each of the 35 CoRoT transits. Practically, we added the following bayesian penalty $BP_{\rm timings}$ to our merit function:
\begin{equation}\label{eq:5}
BP_{\rm timings} = \sum_{i=1,35} \bigg(\frac{T_0 + N_i \times P - T_i}{\sigma_{T_i}} \bigg)^2
 \end{equation}\noindent where $T_i$ is the transit timing determined by Bean (2009) for the $i^{th}$ CoRoT transit, $\sigma_{T_i}$ is its error and $N_i$ is its differential epoch compared to our VLT transit. This procedure relies on the reasonable assumption that the timings determined by Bean  (2009) are uncorrelated with the other transit parameters.
 
 \subsection{Photometric correlated noise and RV jitter noise}
 
Our analysis was done in 4 steps. First, a single MCMC chain was performed. This chain was composed of 10$^6$ steps, the first 20\% of each chain being considered as its burn-in phase and discarded.   The best-fitting model found in the first chain was used to estimate the level of correlated noise in each photometric time-series and a jitter noise in the RV time series. For both photometric time-series,  the red noise was estimated as described in Gillon et al. (2006), by comparing the $rms$ of the unbinned and binned residuals.  We used a bin size corresponding to a duration of 20 minutes, similar to the timescale of the ingress/egress of the transit. The obtained results were compatible with a purely 
Gaussian noise for both time-series. Still, it is possible that a low-amplitude correlated noise damaging only the eclipse part had been `swallowed'  by our best-fitting model, so we preferred to be conservative and to add quadratically a red noise of 100 ppm to the theoretical uncertainties of each photometric time-series. The deduced RV jitter noise was high: 23 m.s$^{-1}$. Nevertheless, we noticed that it goes down to zero if we discard  the second RV measurement of the first SOPHIE epoch. Furthermore, this measurement has a significantly larger error bar than the others, we thus decided to consider it as doubtful and to do not use it in our analysis. A theoretical jitter noise of 3.5 m.s$^{-1}$ was then added quadratically to the error bars of the other SOPHIE measurements, a typical value for a quiet solar-type star like CoRoT-1 (Wright 2005).
 
 \subsection{Determination of the stellar density}
 
Then, 10 new MCMC chains were performed using the  updated measurement error bars. These 10 chains were then combined, using the  Gelman and Rubin statistic (Gelman \& Rubin 1992) to verify that they were well converged and mixed enough, then the best-fitting values and error bars for each parameter were obtained from their distribution. The goal of this  MCMC run was to provide us with an improved estimation of the stellar density $\rho_\ast $ (see e.g. Torres 2007). The stellar density that we obtained was $\rho_\ast = 0.84^{+0.11}_{-0.07}$ $\rho_{\odot}$. 

\subsection{Stellar-evolutionary modeling} 
The deduced stellar density and the spectroscopic parameters were then used to better constrain the
stellar mass and age via a comparison with theoretical stellar evolution models.Two
independent stellar analysis were performed in order to assess the impact of the
stellar evolution models used on the final system parameters:

\begin{itemize}

\item Our first analysis was based on Girardi's evolution models (Girardi et al.\
2000).  We first perform a linear interpolation between the solar (Z=0.019) and subsolar
(Z=0.008) metallicity theoretical models to derive a set of mass tracks at the metallicity of the
host star ([M/H]=-0.3).  We then compare the effective temperature and the inverse cube root of
the stellar density to the same values in the host star metallicity models.  We interpolate
linearly along the mass tracks to generate an equal number of age points between the zero age
main sequence and the point corresponding to core hydrogen exhaustion.  We then interpolate
between the tracks along equivalent evolutionary points to find the mass, $M=0.94$ $M_{\odot}$, 
and age, $\tau$ = 7.1 ~Gyr, of the host star that best match the measured temperature and stellar density.  
We repeat the above prescription using the extreme values of the observed parameters to determine the uncertainties on the derived mass and age.  The large errors on the spectroscopic parameters, particularly the $\pm 0.25$~dex uncertainty on the metallicity, lead to a 15-20 \% error on the stellar mass ($M=0.94^{+0.19}_{-0.16} M_{\odot}$)
 and an age for the system no more precise  than older than 0.5~Gyr.  Fig. 3 presents the deduced position of CoRoT-1 in a $R/M^{1/3}$-$T_{eff}$ diagram.

\item In the second analysis we apply the Levenberg-Marquard miniminization algorithm to derive the fundamental parameters of the host star. The merit function is defined by:
\begin{equation}
\chi^2=\sum_{i=1}^{3}\,\frac{(O_i^{obs}-O_i^{theo})^2}{(\sigma_i^{obs})^2}
\end{equation}
The observables ($O_i^{obs}$) we take into consideration are effective temperature, surface metallicity and mean density. The corresponding observational errors are $\sigma_i^{obs}$. The theoretical  values ($O_i^{theo}$) are obtained from stellar evolution models computed with the code CLES (Code Li\'egois d'Evolution Stellaire, Scuflaire et al. 2008). Several fittings have been performed, in all of them we use the mixing-length theory (MLT) of convection (B\"ohm-Vitense, 1958) and the most recent equation of state from OPAL (OPAL05, Rogers \& Nayfonov, 2002). Opacity tables are those from OPAL (Iglesias \& Rogers, 1996) for two different solar mixtures, the standard one from Grevesse \& Noels (1993, GN93) and the recently revised solar mixture from Asplund, Grevesse \& Sauval (2005, AGS05). In the first case $(Z/X)_\odot$=0.0245, in the second one $(Z/X)_\odot$=0.0167. These tables are extended at low temperatures with Ferguson et al. (2005) opacity values for the corresponding metal mixtures. The surface boundary conditions are given by grey atmospheres with an Eddington law.  Microscopic diffusion (Thoul et al. 1994) is included in stellar model computation. The parameters of the stellar model are mass, initial hydrogen ($X_{\rm i}$) and metal ($Z_{\rm i}$) mass fractions, age, and the parameters of convection ($\alpha_{\rm MLT}$ and the  overshooting  parameter). Since we have only three observational constraints, we decide to fix the $\alpha_{\rm MLT}$ and $X_{\rm i}$ values to those derived from the solar calibration for the same input physics. Furthermore, given the low mass we expect for the host star, all the models are computed without overshooting.
The values of stellar mass and age obtained for the two different solar mixtures are:
$M=0.90\pm 0.21\,M_{\odot}$ with GN93 and $M=0.92\pm 0.18 M_{\odot}$ with AGS05, and respectively $\tau=7.5\pm6.0$~Gyr  and $\tau=6.9\pm 5.4$~Gyr. 

\end{itemize}

The result of our two independent stellar analysis are thus fully compatible, and the uncertainty due to the large errors on the spectroscopic parameters dominates the one coming from  our imperfect knowledge of  stellar physics. The large uncertainties affecting the stellar mass and age are mainly due to the  lack of accuracy in metallicity determination.  We estimate from several tests that an improvement in the atmospheric parameters determination leading to an error in metallicity of 0.05~dex would translate in a reduction in uncertainty by  a factor three for the stellar mass, and a factor two for the stellar age. Moreover, decreasing the effective temperature error to 75~K would imply a subsequent reduction of stellar parameter errors by an additional factor two. Getting more high-SNR high-resolution spectroscopic data for the host star is thus very desirable.

\subsection{Determination of the system parameters}

For the last part of our analysis, we decided to use 0.93 $\pm$ 0.18 $M_{\odot}$, i.e. the average of the values obtained with the two different evolution models, as our starting value for the stellar mass.
A new MCMC run was then performed. This run was identical to the first one, with the exception that $M_\ast$ was also a jump parameter under the control of a Bayesian penalty based on $M_\ast$ = 0.93 $\pm$ 0.18 $M_\odot$. At each step of the chains, the physical parameters $M_p$, $R_p$ and $R_\ast$ were computed from the relevant jump parameters including the stellar mass. Table 1 shows the deduced values for the jump + physical parameters and compares them to the values presented in B08. It also shows the Bayesian penalties used in this second MCMC run.

\begin{figure}
\label{fig:c}
\centering                     
\includegraphics[width=9cm]{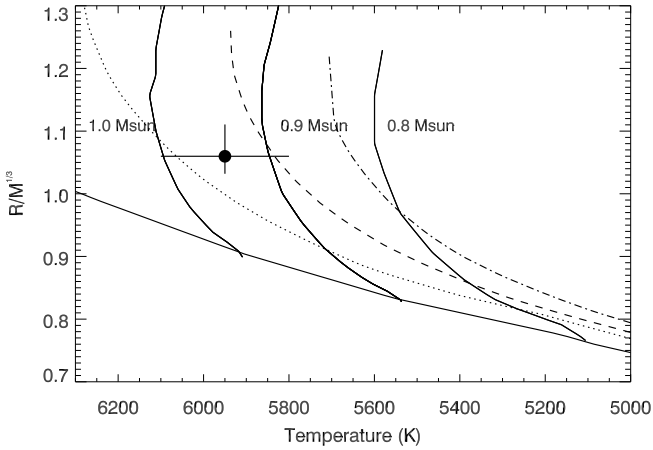}
\caption{$R/M^{1/3}$ (in solar units) versus effective temperature 
for CoRoT-1 compared to the theoretical stellar stellar evolutionary 
models of Girardi et al. (2000) interpolated at -0.3 metallicity.
The labeled mass tracks are for 0.8, 0.9 and 1.0 $M_\odot$ and the 
isochrones are 100 Myr (solid), 5 Gyr (dotted), 10 Gyr (dashed), 16 Gyr (dot-dashed).  
We  have interpolated the tracks at -0.2 metallicity and have included the uncertainty 
on the metallicity ($\pm$0.25) in the overall uncertainties on the mass and the age. }
\end{figure}

\begin{table*}
\label{tab:params}
\begin{tabular}{lcccccl}
\hline
Parameter  & Value & Bayesian penalty & B08 & Unit &\\ \noalign {\smallskip}
\hline \noalign {\smallskip}
$Jump$ $parameters$ & & & & \\ \noalign {\smallskip}
\hline \noalign {\smallskip}
Transit epoch  $ T_0  $ & $ 2454524.62324^{+0.00009}_{-0.00013}$ &  & 2454159.4532 $\pm$ 0.0001& BJD  \\ \noalign {\smallskip}
Planet/star area ratio  $ (R_p/R_s)^2 $ & $ 0.01906^{+0.00020}_{-0.00040} $ &  & $0.01927 \pm 0.00058$ &  \\ \noalign {\smallskip}
Transit width  $W$ & $ 0.10439 \pm 0.00094 $ &  & &  day  \\ \noalign {\smallskip}
2.09 $\mu$m occultation depth & $0.00278^{+ 0.00043}_{- 0.00066}$  & & & \\ \noalign {\smallskip}
$ b'=a\cos{i}/R_\ast $ & $ 0.398^{+ 0.032}_{- 0.043} $  & & $0.420 \pm 0.043$ &  $R_*$  \\ \noalign {\smallskip}
RV $K_2$ & $215^{+15}_{-16}$ & & $216 \pm 13$ &    \\ \noalign {\smallskip}
RV $\gamma_1$ & $23.366^{+0.020}_{-0.017}$ &  & & \kms   \\ \noalign {\smallskip}
RV $\gamma_2$ & $23.350^{+0.012}_{-0.011}$ &  & & \kms  \\ \noalign {\smallskip}
$e\cos{\omega}$ & $0.0083^{+0.0038}_{-0.0025}$ &  &  & \\ \noalign {\smallskip}
$e\sin{\omega}$  & $-0.070^{+0.029}_{-0.042}$ & &  & \\ \noalign {\smallskip}
$A_{\rm transit}$ & $0.99963^{+0.00028}_{-0.00009}$ & & & \\ \noalign {\smallskip}
$B_{\rm transit}$ & $0.017^{+0.003}_{-0.018}$ & & & day$^{-2}$ \\ \noalign {\smallskip}
$C_{\rm transit}$ & $-0.10^{+0.12}_{-0.02}$  & & & day$^{-1}$ \\ \noalign {\smallskip}
$A_{\rm  occultation}$ &$1.00041^{+0.00096}_{-0.00052}$ & & & \\ \noalign {\smallskip}
$B_{\rm occultation}$ &$-0.008^{+0.007}_{-0.023}$ &  &  & day$^{-2}$ \\ \noalign {\smallskip}
$C_{\rm occultation}$ & $0.029^{+0.079}_{-0.029}$& & & day$^{-1}$ \\ \noalign {\smallskip}
& & & &  \\
Orbital period  $ P$ & $ 1.5089686^{+ 0.0000005}_{- 0.0000006} $ & from timings in Bean (2009) & 1.5089557 $\pm$ 0.0000064 &  day  \\ \noalign {\smallskip}
Stellar mass  $ M_\ast $ & $ 1.01^{+0.13}_{-0.22}$ & 0.93 $\pm$ 0.18 &  0.95 $\pm$ 0.15 & $M_\odot$  \\ \noalign {\smallskip}
R-filter $c_1$ & $ 0.794^{+ 0.047}_{- 0.048}$ & 0.909 $\pm$ 0.067 &  &  \\ \noalign {\smallskip}
R-filter $c_2$  & $ -0.444^{+ 0.054 }_{- 0.032}$ & -0.423 $\pm$ 0.046 &   &  \\ \noalign {\smallskip}
\hline \noalign {\smallskip}
$Deduced$ $parameters$      &    &  &  &  \\ \noalign {\smallskip}
\hline \noalign {\smallskip}
RV $K$ & $ 188 \pm 14 $  & & 188 $\pm$ 11& \ms \\ \noalign {\smallskip}
$b_{transit}$ &   $ 0.426^{+ 0.035}_{- 0.042} $  & &  $0.420 \pm 0.043$  &  $R_*$ \\ \noalign {\smallskip}
$b_{occultation}$ &   $ 0.370^{+ 0.037}_{- 0.049} $  & &  $0.420 \pm 0.043$  &  $R_*$ \\ \noalign {\smallskip}
Orbital semi-major axis $ a $ & $ 0.0259 ^{+ 0.0011}_{- 0.0020} $  & & $0.0254 \pm 0.0014$ & AU \\ \noalign {\smallskip}
Orbital inclination  $ i $ & $ 85.66^{+0.62}_{-0.48} $ & & 85.1 $\pm$ 0.5& degree \\ \noalign {\smallskip}
Orbital eccentricity $ e $ & $ 0.071^{+0.042}_{-0.028} $ &  &  0 (fixed)&  \\ \noalign {\smallskip}
Argument of periastron  $ \omega $ & $276.7^{+5.9}_{-4.3}$ &   & &  degree  \\ \noalign {\smallskip}
Stellar radius  $ R_\ast $ & $ 1.057^{+ 0.055}_{- 0.094} $ &  & 1.11 $\pm$ 0.05 &  $R_\odot$ \\ \noalign {\smallskip}
Stellar density  $\rho_* $ & $0.86^{+ 0.13}_{- 0.08} $ & & $0.698 \pm 0.033$ &  $\rho_\odot $\\ \noalign {\smallskip}
R-filter $u_1$ & $ 0.229^{+ 0.025 }_{- 0.022}$ &  & & \\ \noalign {\smallskip}
R-filter $u_2$ & $ 0.336^{+ 0.012}_{- 0.020}$ & & & \\ \noalign {\smallskip}
Planet radius  $ R_p $ & $ 1.45 ^{+ 0.07}_{- 0.13} $ & & 1.49 $\pm$ 0.08 & $R_J$ \\ \noalign {\smallskip}
Planet mass  $ M_p $ & $ 1.07 ^{+ 0.13}_{- 0.18} $  & & 1.03 $\pm$ 0.12 & $M_J$ \\ \noalign {\smallskip}
Planet density  $ \rho_p $ & $0.350^{+0.077}_{-0.042}$ & & $0.31 \pm 0.06$ &  $\rho_{J}$ \\ \noalign {\smallskip}
\hline\\ \noalign {\smallskip}
\end{tabular}
\caption{CoRoT-1 system parameters and 1-$\sigma$ error limits derived from the MCMC analysis. The parameters $A$, $B$ and $C$ are the zero-, first- and second-order coefficients of the polynomial used to model the photometric trend. The values and error bars used in the Bayesian penalties are shown in the third column. Fourth column shows the values  presented in B08. }
\end{table*}

\section{Discussion}

\subsection{The density and eccentricity of CoRoT-1b}

As can be seen in Table 1, the transit parameters that we obtain from our VLT/FORS-2 R-band photometry agree well with the ones presented in B08 and based on CoRoT photometry. Our value for the transit impact parameter is in good agreement with the one obtained by B08, and has a similar uncertainty. The planet/star area ratio that we deduce is within the error bar of the values obtained by B08, while our error bar is smaller. Our deduced physical parameters agree also very well with the ones presented in B08. Our analysis confirms thus the very low-density of the planet (see Fig. 4) and its membership to the sub-group of short period planets too large for current models of irradiated planets (Burrows et al. 2007; Fortney et al. 2008). 

In this context, it is worth noticing the marginal non-zero eccentricity that we deduce from our combined analysis: $e  = 0.071^{+0.042}_{-0.028}$. As outlined by recent works (Jackson et al. 2008b, Ibgui \& Burrows 2009), tidal heating could play a major role in the energy budget of very short period planets and help explaining the very low density of some of them. Better constraining the orbital eccentricity of CoRoT-1b by obtaining more radial velocity measurements and occultation photometry is thus desirable. To test the amplitude of the constraint brought by the occultation on the orbital eccentricity, we made an analysis similar to the one presented in Sec. 3 but discarding the HAWK-I photometry. We obtained similar results for the transit parameters, but the eccentricity was poorly constrained: we obtained much less precise values for $e \cos{\omega}$ and $e \sin{\omega} $, respectively $0.020^{+0.024}_{-0.029}$ and $-0.170^{+0.062}_{-0.078}$. HAWK-I occultation brings thus a strong constraint on these parameters, especially on $e \cos{\omega}$. 

Table 1 shows that our analysis does not agree with B08 for one important parameter: the stellar density. Indeed, the value presented in B08 is significantly smaller and more precise than ours. Still, B08 assumed a zero eccentricity in their analysis, while the stellar mean density deduced from transit observables depend on $e$ and $\omega$ (see e.g. Winn 2009). To test the influence of the  zero eccentricity assumption on the deduced stellar density, we made a new MCMC analysis assuming $e$ = 0. We obtained this time $\rho_\ast  = 0.695^{+0.043}_{-0.030}$ $\rho_\odot$, in excellent agreement with the value $\rho_\ast  = 0.698 \pm 0.033$ $\rho_\odot$ presented by B08. This shows nicely that not only are VLT and CoRoT data fully compatible, but also that assuming a zero eccentricity can lead to an unreliable stellar density value and uncertainty. In our case, this has no significant impact on the deduced physical parameters because the large errors that we have on the stellar effective temperature and metallicity dominate totally the result of the stellar-evolutionary modeling (see Sec. 3.6). Still, the point is important. As shown by Jackson et al. (2008a), 
most published estimates of planetary tidal circularization timescales used inappropriate assumptions that led to unreliable values, and most close-in planets could probably keep a tiny but non-zero eccentricity during a major part of their lifetime. In this context, very precise transit photometry like the CoRoT one is not enough to reach the highest accurary on the physical parameters of the system, a precise determination of $e$ and $\omega$ is also needed. This strengthens the interest of getting complementary occultation photometry in addition to high-precision radial velocities to improve the characterization of transiting planets. 

\begin{figure}
\label{fig:e}
\centering                     
\includegraphics[width=8cm]{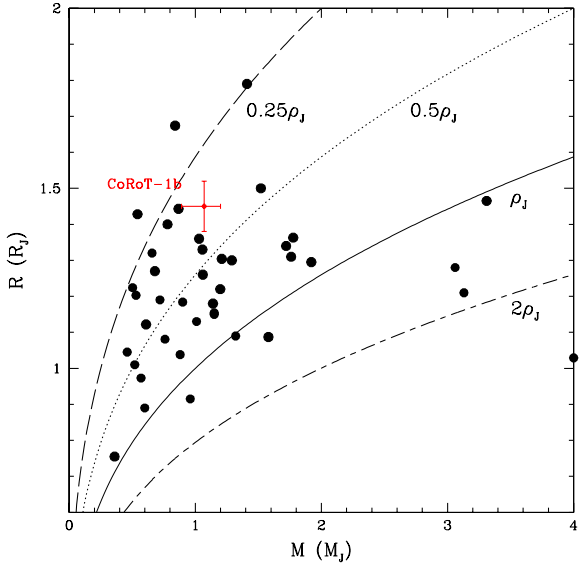}
\caption{Position of CoRoT-1b (in red) among the other transiting planets (black circles, values from http://exoplanet.eu)  in a  mass-radius diagram. The error bars are shown only for CoRoT-1b for the sake of clarity. }
\end{figure}

\subsection{The atmospheric properties of CoRoT-1b}

The  flux at 2.09 $\mu$m of this planet is slightly larger than the (zero-albedo) equilibrium temperature,
$\sim$ 2660 K, obtained if the star's effective temperature is allowed to be as high as 6100K
(maximum within the 1-$\sigma$ error-bars from B08).  An irradiated planet atmosphere
model (following Barman et al. 2005) for CoRoT-1b, was computed adopting the maximum
observational allowed stellar effective temperature and radius and assuming zero energy is
transported to the night side.  Solar metallicity was assumed and all other parameters
were taken from Table 1.  This model (Fig. 5) falls short of matching the observations within 1-$\sigma$,
while a black body with the same equilibrium temperature as the irradiated planet model is in
better agreement.   The atmosphere model is hot enough for a significant temperature inversion
to form for P $<$ 0.1 bar and is nearly isothermal from 0.1 down to $\sim$ 100 bar.   A model which uniformly
redistributes the absorbed stellar flux across the entire planet surface (lower curve in Fig. 5) is far too
cool to match the observations and is excluded at $\sim$ 3 $\sigma$.   The flux at 2.09 $\mu$m alone is suggestive
that very little energy is redistributed to the night side; however additional observations at
shorter and/or longer wavelengths are needed to better estimate the bolometric flux of the planet's
day side.  Occultation measurements in other bands will help provide limits on the
day side bolometric flux and determine the depth of any possible temperature inversion and the extent
of the isothermal zone.

Recently Snellen et al. (2009) measured the dayside planet-star flux ratio of CoRoT-1
in the optical ($\sim$ 0.7 $\mu$m) to be $1.26 \pm 0.33 \times 10^{-4}$.   The hot, dayside only, model
shown in Fig. 5 predicts a value of $1.29 \pm 0.33 \times 10^{-4}$, which is fully consistent with the
optical measurement.   Consequently, it appears as though very little energy is
being carried over to the night side of this planet.

\begin{figure}
\label{fig:e}
\centering                     
\includegraphics[width=9cm]{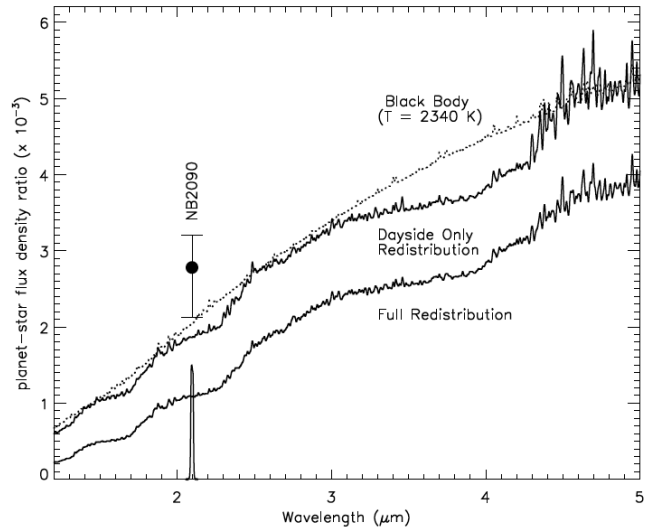}
\caption{Comparison of our 2.09 $\mu$m occultation depth measured for CoRoT-1 with models of planet-star flux density ratios assuming that the absorbed stellar flux is redistributed across the dayside only (top curve) and uniformly redistributed across the entire planetary atmosphere (lower curve).  A black body model is also shown (dotted) for T = 2365 K. }
\end{figure}

\subsection{Assessing the presence of another body in the system}

As shown in Table 1, our deduced systemic RV for each SOPHIE epoch agree with well each other: we 
do not confirm the RV drift mentioned in B08. Our combined analysis presented in Sec. 3  leads to a very precise determination of the orbital period: $ 1.5089686^{+ 0.0000005}_{- 0.0000006} $ days, thanks to a lever arm of nearly one year between CoRoT transits and the VLT one. A simple linear fit for timing $vs$ epoch based on the CoRoT and VLT transits lead to 
a similar level of precision, giving $P=1.5089686^{+0.0000003}_{-0.0000005}$ days. This fit has a reduced $\chi^2$ of 1.28 and  the $rms$ of its residuals (see Fig. 6) is 36~s. These values are fully consistent with those reported by Bean (2009) for CoRoT data alone. We also notice the same 3-$\sigma$ discrepancy with transit \#23 which, once removed, results in a reduced $\chi^2$ of 1.00, hereby confirming the remarkable periodicity of the transit signal. 

\begin{figure}
\label{timings_corot1b}
\centering
\includegraphics[width=8.0cm]{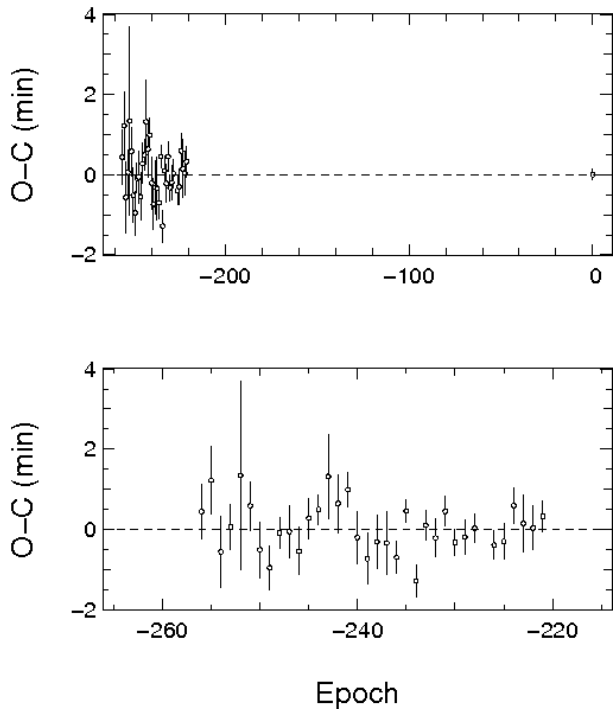}
\caption{$Top$: Residuals of the linear fit timing $vs$ epoch for CoRoT-1b (see text for details). The rightmost point is our VLT/FORS2 timing. $Botttom$: zoom on the CoRoT residuals.}
\end{figure}

While limits on additional planetary companions in CoRoT-1 system have been extensively discussed for transit timing variations (TTVs) by Bean (2009), it is interesting to take limits from RVs into account. Figure 7 illustrates the domains where additional planets could be found through TTVs (white) and through RV measurements (above coloured curves). We focused on short periods objects since TTVs are more sensitive to nearby perturbators as compared to the known transiting planet. We assumed an eccentricity of 0.05 for a putative coplanar planet and used the $Mercury$ package described in Chambers (1999) to estimate by numerical integration the maximum TTV signal expected for CoRoT-1b. White is the domain with a $>$ 5-$\sigma$ detection through TTVs according to CoRoT data $rms$, while black area is below the 1-$\sigma$ detection threshold. Although approximative, this shows that for a typical 3 \ms  ~accuracy on radial velocities (dashed curve on Fig. 7) routinely obtained with HARPS spectrograph (Mayor et al. 2003), no room remains for planetary companion detection through TTVs alone. Thus, TTV search method may opportunely be applied on active and/or stars for which RV measurements accuracy is limited, increasing a detectability area for which RVs are not or far less sensitive to. 

Each transit timing may be compared to  a single RV measurement. The increased free parameters in a TTV orbital solution raise degeneracies that could not be lifted by considering the same number of datapoints that would allow an orbital solution recovery with RVs. Determination of a large number of consecutive transits added to occultation timings allow to access to the uniqueness of the solution as well as lowering constraints on timings accuracy (Nesvorn\'y \& Morbidelli 2008). This is thus a high cost approach that is the most potentially rewarding on carefully determined targets stars.

\begin{figure}
\centering
\includegraphics[width=7.0cm]{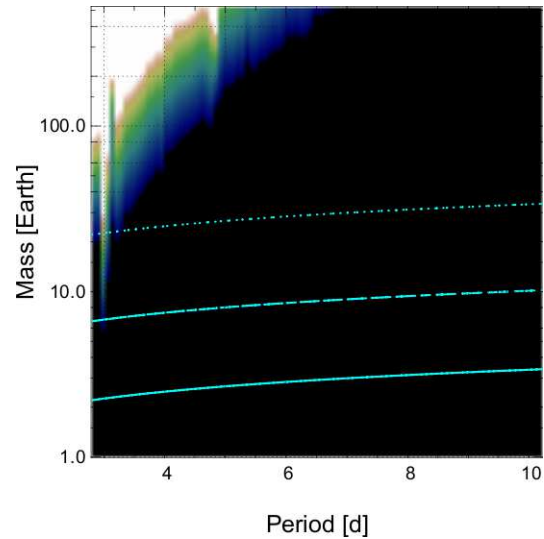}
\caption{Detectivity domain for a putative CoRoT-1c planet, assuming $e_c=0.05$. In white, the period-mass region where planets yield maximum TTV on CoRoT-1b above 100 s ($5\sigma$ detection based on CoRoT data). Companions in the black area yield maximum TTV below the $1\sigma$ threshold. Solid, dashed and dotted curves shows RV detection limits for 1, 3 and 10 m/s RMS. }
\end{figure}

\section{Conclusion}

We have obtained new high-precision transit photometry for the planet CoRoT-1b. Our deduced system parameters are in very good agreement with the ones presented in B08, providing thus an independent verification of the validity of the CoRoT photometry. Due to the precision of the CoRoT and VLT transit photometry and the long baseline between them, the orbital period is now known to a precision better than 1/10th of a second. The precision on the planetary mass and radius is limited by the large errors on the stellar spectroscopic parameters, and a significant precision improvement should be made possible by getting new high-quality spectra of CoRoT-1.

We have also measured successfully the occultation of the planet with HAWK-I, a new wide-field near-infrared imager mounted recently on the VLT. The large occultation depth that we measure is better reproduced by an atmospheric model with no redistribution of the absorbed stellar flux to the night side of the planet. This measurement firmly establishes the potential of the HAWK-I instrument for the study of exoplanetary atmospheres. At the time of writing, $Spitzer$ cryogen is nearly depleted and soon only its 3.6 $\mu$m and 4.5 $\mu$m will remain available for occultation measurements, while the eagerly awaited JWST (Gardner et al. 2006) is not scheduled for launch before 2013. It is thus reassuring to note that ground-based near-infrared photometry is now able to perform precise planetary occultation measurements, bringing new independent constraints on the orbital eccentricity and on the atmospheric physics and composition of highly irradiated extrasolar planets.

\begin{acknowledgements} 
The authors thank  the VLT staff for its support during the preparation and acquisition of the observations. In particular, J. Smoker and F. Selman are gratefully acknowledged for their help and support during the HAWK-I  run. C. Moutou and F. Pont are acknowledged for their preparation of the FORS2 observations. F. Bouchy, PI of the ESO Program 080.C-0661(B), is also gratefully acknowledged. M. Gillon acknowledges support from the Belgian Science Policy Office in the form of a Return Grant, and thanks A. Collier Cameron for his help during the development of his MCMC analysis method. J. Montalb\'an thanks A. Miglio for his implementation of the optimization algorithm used in her stellar-evolutionary modeling. We thank the referee Scott Gaudi for a critical and constructive report.
\end{acknowledgements} 

\bibliographystyle{aa}
{}
\end{document}